\documentclass[letterpaper, 10pt, conference]{ieeeconf}

\IEEEoverridecommandlockouts
\makeatletter
\let\NAT@parse\undefined
\makeatother

\usepackage{graphicx}
\usepackage{amsmath}
\usepackage{amssymb}
\usepackage{float}
\usepackage{hyperref}
\hypersetup{colorlinks=true, linkcolor=black, citecolor=blue, urlcolor=black}

\title{\LARGE \bf
Calibrate Once, Fly Any Team: Residual-Grounded Low-Fidelity Training for Cooperative Drone Swarms
}

\author{Maxim Mednikov$^{1}$ and Oren Gal$^{2}$
\thanks{$^{1}$Swarm \& AI Lab (SAIL), University of Haifa, Israel
        {\tt\small mmedniko@campus.haifa.ac.il}}%
\thanks{$^{2}$Swarm \& AI Lab (SAIL), University of Haifa, Israel
        {\tt\small orengal@univ.haifa.ac.il}}%
}

\begin{document}

\maketitle
\thispagestyle{empty}
\pagestyle{empty}

\begin{abstract}
Training multi-agent drone-swarm policies directly in high-fidelity (HF) rigid-body physics is accurate but computationally expensive. This cost scales poorly with team size, as each additional agent multiplies contact-resolution complexity and sharply raises the in-simulation crash rate. To address this, we propose a mixed-fidelity training scheme that eliminates HF reinforcement learning entirely.

A single shared, decentralized policy is optimized inside a fully-differentiable, JAX-native low-fidelity (LF) point-mass simulator. The simulator is corrected by a small, per-agent bagged residual ensemble fit once, offline, using short calibration flights in the HF simulator. Because calibration requires only one isolated drone, the data collection budget does not compound with team size. Reference trajectories are generated by rolling out an existing LF-only policy and tracked in the HF simulator by a zero-training PD controller.

Evaluated across four cooperative drone tasks and team sizes from 3 to 18, the residual-corrected policy outperforms an uncorrected LF baseline in all combinations, and a from-scratch HF policy in 22 of 24 combinations tested. It trails an HF-finetuned policy by a margin that narrows steadily with team size. Ultimately, the proposed method achieves near-equivalent performance at the largest team sizes at a fraction of the computational cost, completely avoiding the high crash rates typical of HF training.
\end{abstract}

\section{Introduction}


Reinforcement learning for robotic control faces a persistent trade-off between simulator speed
and simulation fidelity \cite{collinsReviewPhysicsSimulators2021}. A low-fidelity (LF) point-mass
simulator parallelizes massively and iterates quickly, but its dynamics diverge from a real rigid
body: rotational inertia, momentum, and attitude dynamics introduce a delay between a commanded
attitude change and the resulting displacement that a point-mass model does not exhibit. A policy
trained only on LF physics routinely over-corrects and fails to transfer, an instance of the
broader sim-to-real reality gap \cite{aljalboutRealityGapRobotics2025}.

Training directly in high-fidelity (HF) rigid-body physics does not scale well to multi-agent
settings. Every additional HF agent multiplies the cost of resolving rigid-body contacts, and a
joint $N$-agent rollout must also resolve genuine inter-agent collisions; in our own experiments
the in-simulation crash rate climbs sharply as team size grows, purely from running reinforcement
learning directly in HF physics, and it previews the cost and risk such training would ultimately
incur on physical vehicles. Prior work on multi-robot transfer instead trains inside a deliberately
abstract or reduced-fidelity simulator and reintroduces real-world detail only at deployment
\cite{labiosaMultiRobotCollaborationReinforcement2025,nooraniAbstractionRealityDARPAs2025,samakMixedRealityDigitalTwins2025};
we instead keep a single differentiable LF simulator throughout optimization and correct it
in-place with a learned per-agent residual, rather than switching representations between training
and deployment.

This work asks whether the fidelity gap can instead be closed with a small learned correction
applied to the cheap simulator, fit once, offline, from a short calibration rollout in the HF
simulator. The recipe has three steps. First, a zero-training PD controller flies the HF simulator
along a reference trajectory that itself comes from cheaply rolling out an existing LF-only policy;
any adequate trajectory follower would serve here, and nothing downstream depends on the choice.
Second, a small per-agent residual acceleration model (a bootstrap-bagged ensemble of MLPs) is fit
once on that calibration data and then frozen. Third, a shared multi-agent policy is optimized
entirely inside the LF simulator, corrected by the frozen residual at every step, via
Back-Propagation Through Time (BPTT)
\cite{moraPODSPolicyOptimization2021,xuAcceleratedPolicyLearning2022,zhangBackNewtonsLaws2024};
this is the only stage that dominates total training cost, and it never runs HF physics at all.

Because the residual's features are strictly per-agent, with no information about team size or
teammates, its calibration data is gathered from one isolated drone at a time. For each team size we
regenerate the reference trajectories and re-fit the residual and shared policy from that
single-drone data. No stage of the pipeline ever flies two drones together, so calibration carries
essentially no crash risk. These collection flights can also run in parallel across 
separated drones that they never share airspace, cutting wall-clock collection time further.

In summary, this paper contributes: (i) a training pipeline that grounds a cheap, differentiable
LF simulator in HF rigid-body dynamics with a per-agent residual, fit once offline from a
zero-training controller's calibration rollouts and then frozen, removing HF reinforcement
learning from multi-agent policy training entirely; (ii) a strictly per-agent residual paired
with a fully-shared policy, so grounding a team of any size needs only round-robin single-drone
calibration flights, never a joint multi-agent rollout, and policy-training cost stays flat in
team size; and (iii) an evaluation across four cooperative tasks and every $N\in\{3,5,7,9,12,18\}$
in which the resulting policy beats an uncorrected LF baseline in all 24 combinations and a
from-scratch HF baseline in 22 of 24, with its lead over the latter widening with team size, and
closes the gap to an HF fine-tune, the strongest baseline, steadily as teams grow, to under
$1\%$ at the largest sizes, all at a small fraction of the HF flight time and training crashes
both HF-trained baselines incur. All results are in simulation; physical-drone validation
is future work.

\section{Related Work}

\subsection{Sim-to-real transfer.} 
Conventional approaches to closing the reality gap train entirely in simulation via domain randomization over dynamics and appearance parameters \cite{openaiLearningDexterousInHand2019,tanSimtoRealLearningAgile2018,pengSimtoRealTransferRobotic2018,hwangboLearningAgileDynamic2019}, or by correcting the simulator's internal parameters to match observed real-world behavior \cite{huangWhatWentWrong2023,aljalboutRealityGapRobotics2025}. 
These methods typically target a single, fixed-fidelity simulator. 
In contrast, we address the fidelity gap between two simulators of different computational costs using a small learned residual, rather than tuning the LF simulator's native parameters.

\subsection{Differentiable simulation and BPTT.}
\label{sec:diff-sim-bptt}
Policy optimization via Backpropagation-Through-Time (BPTT) in a differentiable
simulator avoids the high variance and sample-inefficiency of stochastic
policy-gradient methods, provided the environment exposes gradients end-to-end
\cite{moraPODSPolicyOptimization2021,xuAcceleratedPolicyLearning2022}. For a
$T$-step rollout with $a_t = \pi_\theta(s_t)$, $s_{t+1} = f(s_t, a_t)$,
$s_0\sim\rho_0$, and $J(\theta) = \sum_{t=0}^{T-1} r(s_t, a_t)$, a differentiable
step $f$ yields the policy gradient in closed form:
\begin{equation}
\resizebox{\columnwidth}{!}{$\displaystyle
\nabla_\theta J(\theta) = \sum_{t=0}^{T-1}\bigg(\sum_{i=0}^{t-1}\frac{\partial r_t}{\partial s_t}
\Big(\prod_{j=i+1}^{t-1}\frac{\partial s_{j+1}}{\partial s_j}\Big)\frac{\partial s_{i+1}}{\partial \theta}
+ \frac{\partial r_t}{\partial a_t}\frac{\partial a_t}{\partial \theta}\bigg),
$}
\end{equation}
where $\partial s_{i+1}/\partial\theta$ and $\partial a_t/\partial\theta$ are the
policy's direct parameter sensitivities and the product term is the state-transition
Jacobian chain carrying them forward to step $t$. BPTT evaluates this sum by
reverse-mode automatic differentiation, without sampled return estimates, before a
standard gradient update (e.g., Adam) $\theta_{k+1} = \theta_k + \alpha\nabla_\theta J(\theta_k)$.

This approach has been applied to single-quadrotor control
\cite{heegLearningQuadrotorControl2025a} and to swarm navigation in a differentiable
point-mass simulator \cite{zhangBackNewtonsLaws2024}. We build on this foundation
but add a frozen, offline-fit dynamics correction for cooperative multi-agent tasks,
rather than deploying the uncorrected policy.

For the swarm architecture we use full parameter sharing
\cite{terryRevisitingParameterSharing2023} rather than decentralized communication
\cite{batraDecentralizedControlQuadrotor2021a}, graph networks
\cite{tolstayaLearningDecentralizedControllers2017}, or per-agent low-rank adaptation
\cite{zhangLowRankAgentSpecificAdaptation2025}: our formulation is homogeneous
(every agent's residual and policy inputs are structurally identical), so one shared
policy suffices and preserves the scalability of single-drone calibration
(Section~\ref{sec:generalization}).

\subsection{Residual dynamics correction and online adaptation.}
Learning a residual correction over an imperfect nominal model, rather than replacing
it, is an established approach in robot learning \cite{silverResidualPolicyLearning2019},
aligned with model-based RL methods that combine learned dynamics with analytical
models via uncertainty-aware ensembles for sample-efficient planning
\cite{chuaDeepReinforcementLearning2018}. Such learned models often suffer
from compounding errors over long rollouts \cite{lambertInvestigatingCompoundingPrediction2022},
a failure mode analogous to the closed-loop instability addressed in our residual
design (Section~\ref{sec:stability}).

Recent quadrotor work addresses the sim-to-real gap with online, frequently
updated residual models \cite{huangDATTDeepAdaptive2023,oconnellNeuralFlyEnablesRapid2022}.
\textit{learning\_on\_the\_fly} \cite{panLearningFlyRapid2026} fits an online
residual ensemble for a single quadrotor and is the architectural basis for our
residual-acceleration formulation. Whereas some swarm work learns residual
\emph{interaction} forces between members such as downwash \cite{shiNeuralSwarm2PlanningControl2021a},
our residual stays strictly per-agent, using no information about teammates or team size
(Section~\ref{sec:generalization}).

Unlike continuous online refitting against the active policy, we fit a small bagged
ensemble \cite{panLearningFlyRapid2026} once offline and freeze it, trading an online
loop's tighter distribution matching for a pipeline that runs the HF simulator exactly
once per team size. Concurrent work integrates residual models with BPTT for real-world
quadrotor adaptation \cite{renLearningAgileQuadrotor2026} or conditions policy and
residual on online-inferred latent dynamics \cite{xingContinualRobotPolicy2026}; ours
is calibrated entirely in HF simulation, not on hardware, and uses no multi-agent features.

Closest is \textit{Sym2Real} \cite{leeSym2RealSymbolicDynamics2026}, which builds a
base model in a cheap LF simulator by symbolic regression and adapts it with a learned
residual from roughly ten real trajectories. It targets single-agent LF-to-real
transfer; we extend the paradigm to cooperative swarms, grounding teams of any scale
with a per-agent residual calibrated one drone at a time, never a joint multi-drone HF flight.

\subsection{PD Velocity Control of Quadrotors}
\label{sec:pd-control}
For a quadrotor with fixed yaw, any twice-differentiable position trajectory determines
the required thrust magnitude and attitude in closed form
\cite{mellingerMinimumSnapTrajectory2011}. We use this property with a standard
PD controller to track trajectories and collect calibration data in the HF simulator.
Given a current state $(\mathbf{p},\mathbf{v})$ and a target
$(\mathbf{p}^*,\mathbf{v}^*,\mathbf{a}^*)$, the controller computes a desired thrust 
vector incorporating feedforward acceleration and gravity compensation:
\begin{equation}
\mathbf{T} = \mathbf{a}^* + k_p(\mathbf{p}^*-\mathbf{p}) + k_d(\mathbf{v}^*-\mathbf{v}) + [0,0,g]^\top.
\end{equation}
This thrust vector is then mapped directly into the vehicle's body frame via inverse 
dynamics to yield the roll, pitch, and thrust commands:
\begin{equation}
\resizebox{\columnwidth}{!}{$\displaystyle
\phi_{cmd} = -\arcsin\!\left(\tfrac{T_y}{\lVert\mathbf{T}\rVert}\right),\quad
\theta_{cmd} = \operatorname{atan2}(T_x,T_z),\quad
T_{cmd} = m\lVert\mathbf{T}\rVert.
$}
\end{equation}
This deterministic, zero-training controller is used only to fly the calibration
data (Section~\ref{sec:cheap-pipeline}); any sufficiently accurate trajectory follower
would serve without altering the downstream pipeline.

\begin{figure*}[h]
\centering
\includegraphics[width=\textwidth]{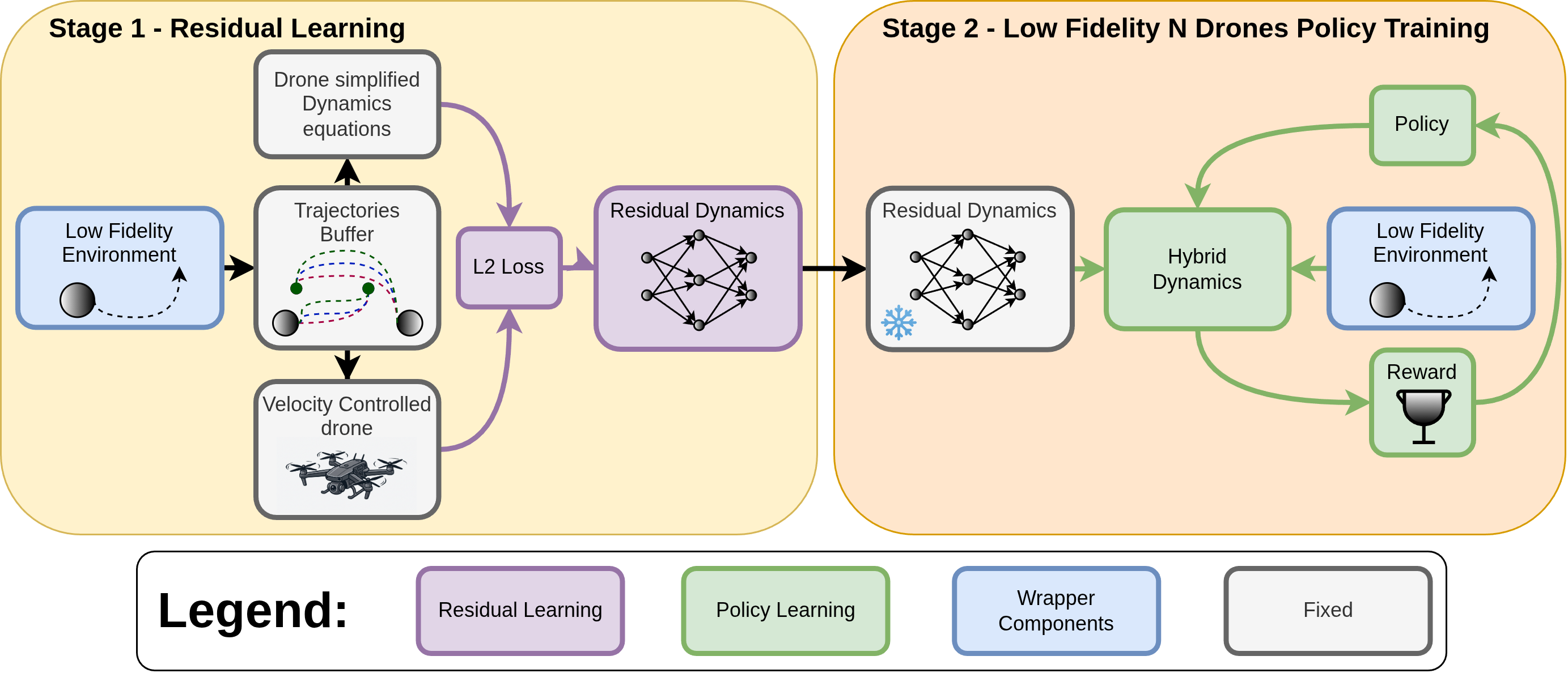}
\caption{Overview of the training pipeline (Section~\ref{sec:cheap-pipeline}).
\emph{Stage 1 (left):} an existing LF-only policy is rolled out in the LF simulator to generate a
reference trajectory, which is tracked in the HF simulator by the PD controller
(Section~\ref{sec:pd-control}); the residual ensemble is then fit, via an L2 regression loss, against
the gap between that HF flight and the LF simulator's own (simplified) dynamics equations.
\emph{Stage 2 (right):} the fitted residual is frozen and composed with the LF simulator into a
corrected dynamics model, inside which a freshly-initialized policy is trained end-to-end via BPTT,
with no further HF-simulation contact.}
\label{fig:scheme}
\end{figure*}

\section{Methodology}
\label{sec:methodology}

The proposed pipeline (Figure~\ref{fig:scheme}) consists of a shared policy optimized entirely 
within LF physics, dynamically corrected at every integration step by a per-agent residual. 
This section details the network architectures and the four-stage training pipeline.

\subsection{Policy and Residual}

Both networks are multi-layer perceptrons with two hidden layers of width $128$. The policy
has a $\tanh$ output layer; the residual is a five-member bootstrap-bagged ensemble whose
per-member $\tanh$ output bounds the correction to $\pm 35\,\mathrm{m/s}^2$.

\subsubsection{Policy}
\label{sec:policy}
The shared policy is applied independently to each agent's observation vector. It is optimized
directly in LF physics via backpropagation-through-time (Section~\ref{sec:diff-sim-bptt}): the
full trajectory rollout is differentiated end-to-end with respect to the policy weights,
requiring neither a critic nor an advantage estimator.

\subsubsection{Residual}
\label{sec:residual}
Each member is independently initialized and trained on its own bootstrap resample of the
calibration transitions; at deployment the member outputs are plain-averaged, with no
disagreement gating or uncertainty weighting. From a 12-dimensional per-agent feature
vector (position $\mathbf{p}$, velocity $\mathbf{v}$, attitude $(\phi,\theta,\psi)$, and
nominal point-mass acceleration $\mathbf{a}$), the ensemble predicts an additive corrective
acceleration $\Delta\mathbf{a}=[\Delta a_x, \Delta a_y, \Delta a_z]$. Because this correction
depends only on an agent's own state and action, calibration data can come from a single
isolated drone (Section~\ref{sec:generalization}).

Two choices keep the residual well-behaved on out-of-distribution states. First, it is fit
\emph{once, offline, and frozen}, removing the moving-target instability of online adaptation.
Second, its output is bounded: because the correction enters the LF integration additively, an
unbounded residual evaluated off-distribution can act as an unstable linear map that diverges to
floating-point overflow. A saturating $\tanh$ output structurally prevents this, with no need
for artificial episode resets or severed gradient paths.\label{sec:stability}

Finally, in tasks with extended station-keeping, near-hover states are under-represented in the
dynamic calibration flights (Section~\ref{sec:experiments}); a per-sample loss reweighting
upweights low-velocity transitions during the fit to compensate.

\subsection{Training Pipeline}
\label{sec:cheap-pipeline}

The residual is fit and the policy trained in four decoupled stages, once per scenario and team
size (Figure~\ref{fig:scheme}):
\begin{enumerate}
    \item \textbf{Generate reference trajectories.} An existing, uncorrected LF-only policy
    (the baseline of Section~\ref{sec:experiments}) is rolled out in the LF simulator. Positions
    and velocities are logged every step, and accelerations reconstructed by exact
    finite-differencing of velocity to match the LF simulator's native integration.
    \item \textbf{Fly the reference in the HF simulator.} Each trajectory in the pool drives an
    isolated drone rather than a joint $N$-agent rollout: trajectory $i$ exercises role
    $i \bmod N$ in a deterministic round-robin schedule, covering all roles without multi-drone
    flight. The PD controller (Section~\ref{sec:pd-control}) tracks it in the HF simulator, the
    only stage that runs HF physics, logging per-step state, action, and next state. With a
    fixed number of trajectories per role, the calibration budget scales linearly with $N$
    (Section~\ref{sec:traj-sweep} analyzes subsampling it).
    \item \textbf{Fit the residual.} For each transition, the regression target is the
    acceleration offset that, added to the nominal LF step, reproduces the observed HF next state
    exactly. The ensemble is fit to these targets once, offline, by full-batch supervised
    regression (Section~\ref{sec:residual}), using no reward signal or policy gradient, and
    frozen thereafter.
    \item \textbf{Train the policy.} A freshly initialized shared policy is optimized by BPTT
    entirely in the LF simulator. The frozen correction is added to each agent's acceleration
    every step as a stop-gradient constant, so the BPTT gradient flows only through the nominal
    point-mass dynamics, never the residual network's input Jacobian. An ensemble trained on a
    narrow calibration manifold has arbitrary off-distribution Jacobians, so backpropagating
    through it would hand the policy spurious gradient directions and destabilize optimization;
    severing that path retains the forward physical correction without injecting gradient noise.
    No HF simulation occurs here.
\end{enumerate}
The resulting policy is then deployed in the HF simulator for evaluation
(Section~\ref{sec:experiments}).

Stage 1 uses an existing LF-only policy rather than an idealized geometric target: an LF policy
commands more aggressive maneuvers than the vehicle can realize, so replaying its actions
open-loop would exceed the HF flight envelope. Tracking its \emph{achieved} path in closed loop
keeps the calibration data dynamically feasible, without HF reinforcement learning.

\subsection{Single-Drone Calibration at Any Team Size}
\label{sec:generalization}
The pipeline scales by eliminating multi-drone HF flights entirely. Because the residual uses
only per-agent features, HF calibration data comes from one isolated drone at a time (Stage 2),
and simulating $N$ LF agents reduces to $N$ independent point-mass updates under the residual,
so policy-optimization cost is largely invariant to team size.

We calibrate the residual and train the policy separately for each $N \in \{3,5,7,9,12,18\}$. The
only cost that scales with $N$ is single-drone calibration collection, driven by the need to
maintain uniform role coverage in the round-robin trajectory pool, not by the sample complexity
of multi-agent interaction. Every tested swarm size is thus grounded and trained without a single
multi-agent HF flight.

\section{Experiments}
\label{sec:experiments}

\subsection{Environments}
Both simulation fidelities share a $[-1,1]^3$ action space, a $dt=0.02$\,s RL
timestep, and a $0.15$\,m agent collision radius.

\subsubsection{Low-Fidelity Model (pmDrones)}
The LF environment is pmDrones, a JAX-native \cite{JaxmlJax2026}, fully 
JIT/vmap-parallelized point-mass simulator. Commanded roll and pitch track the 
target angles via a first-order lag at every simulation step. Thrust tracks its 
command instantaneously, is remapped to a non-negative range, and is converted to 
acceleration via standard attitude-to-thrust-vector decomposition on the post-lag 
angles (with yaw locked at $\psi=0$):
\begin{equation}
\begin{aligned}
a_x &= \tfrac{T}{m}\sin\theta\cos\phi, \\
a_y &= -\tfrac{T}{m}\sin\phi, \\
a_z &= \tfrac{T}{m}\cos\theta\cos\phi - g
\end{aligned}
\end{equation}
The learned residual correction $\Delta\mathbf{a}$ (Section~\ref{sec:residual}) is
added to $\mathbf{a}$ before integration, such that $\Delta\mathbf{a}=\mathbf{0}$
recovers the nominal pmDrones physics. Unlike high-fidelity multirotors, this 
nominal point-mass model completely omits body rotational inertia, rotor gyroscopic 
effects, and velocity-dependent aerodynamic drag. Consequently, the residual ensemble 
must independently capture these structural discrepancy terms to prevent simulated 
trajectories from diverging upon transfer to the full rigid-body evaluation platform.

\subsubsection{High-Fidelity Model (crazyFlow)}
The HF environment is crazyFlow \cite{schuckCrazyflowAccurateGPUAccelerated2026},
a MuJoCo/mujoco-mjx-based, GPU-accelerated, differentiable rigid-body simulator
driven by the same $[\phi_{cmd},\theta_{cmd},T_{cmd}]$ convention. It models aerodynamic
effects such as rotor drag that the point-mass model omits, and is system-identified
against physical Crazyflie hardware rather than hand-tuned, reportedly reaching
sub-centimeter tracking without domain randomization; this makes it an accurate proxy for
real-world deployment. We treat it strictly as ground truth: it is never used inside the
policy-optimization loop, only to generate calibration data (Section~\ref{sec:cheap-pipeline})
and evaluate deployed policies, so all HF results here are in simulation, with hardware
deployment left to future work.

\subsection{Tasks}

\begin{figure}[h]
\centering
\includegraphics[width=\columnwidth]{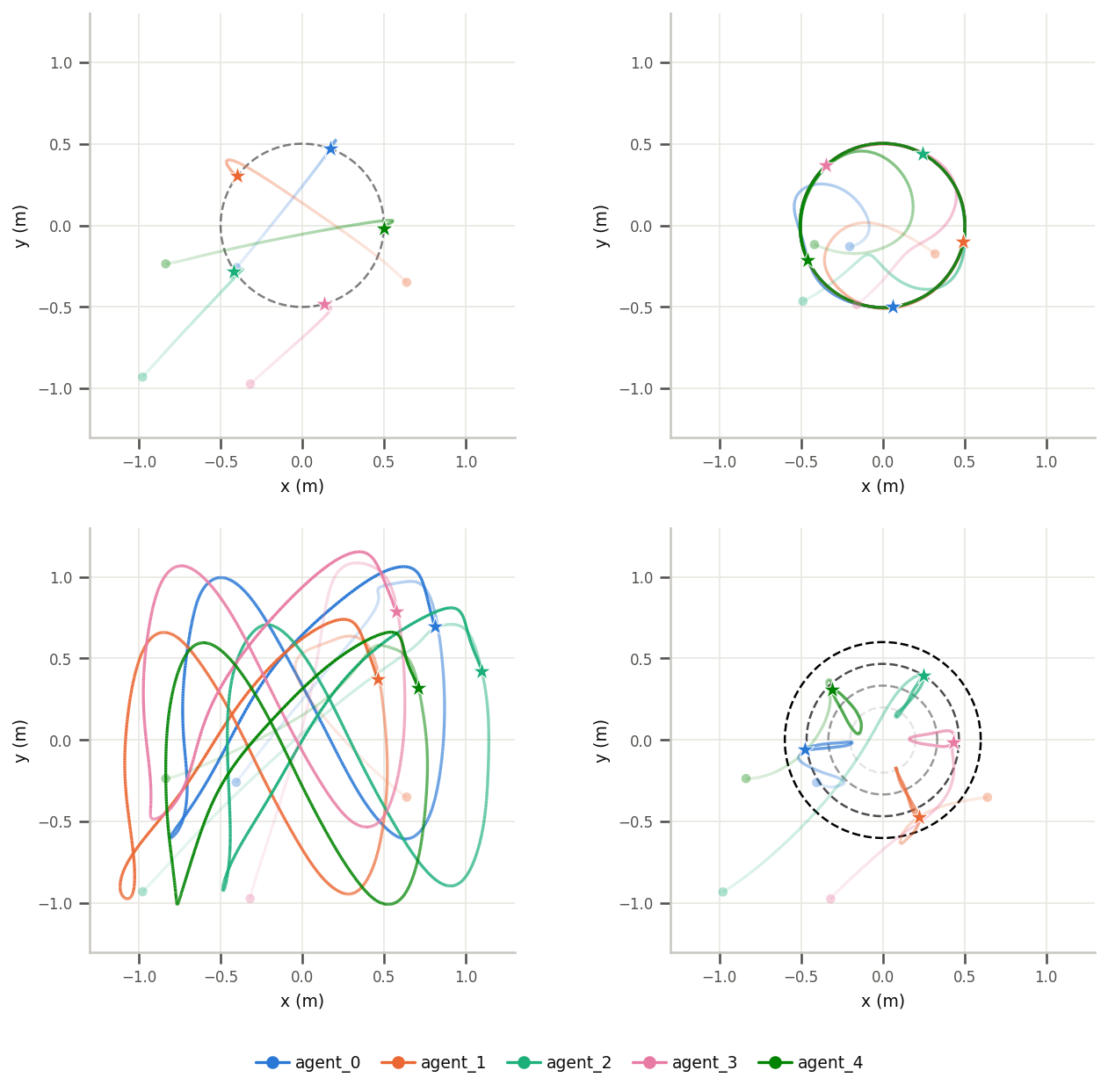}
\caption{Top-view flight trajectories for each of the four tasks, one representative episode at
$N{=}5$ (older positions faded). Circular markers mark each agent's start; stars mark its end.}
\label{fig:scenario_trajectories}
\end{figure}

\begin{figure*}[t]
\centering
\includegraphics[width=\textwidth]{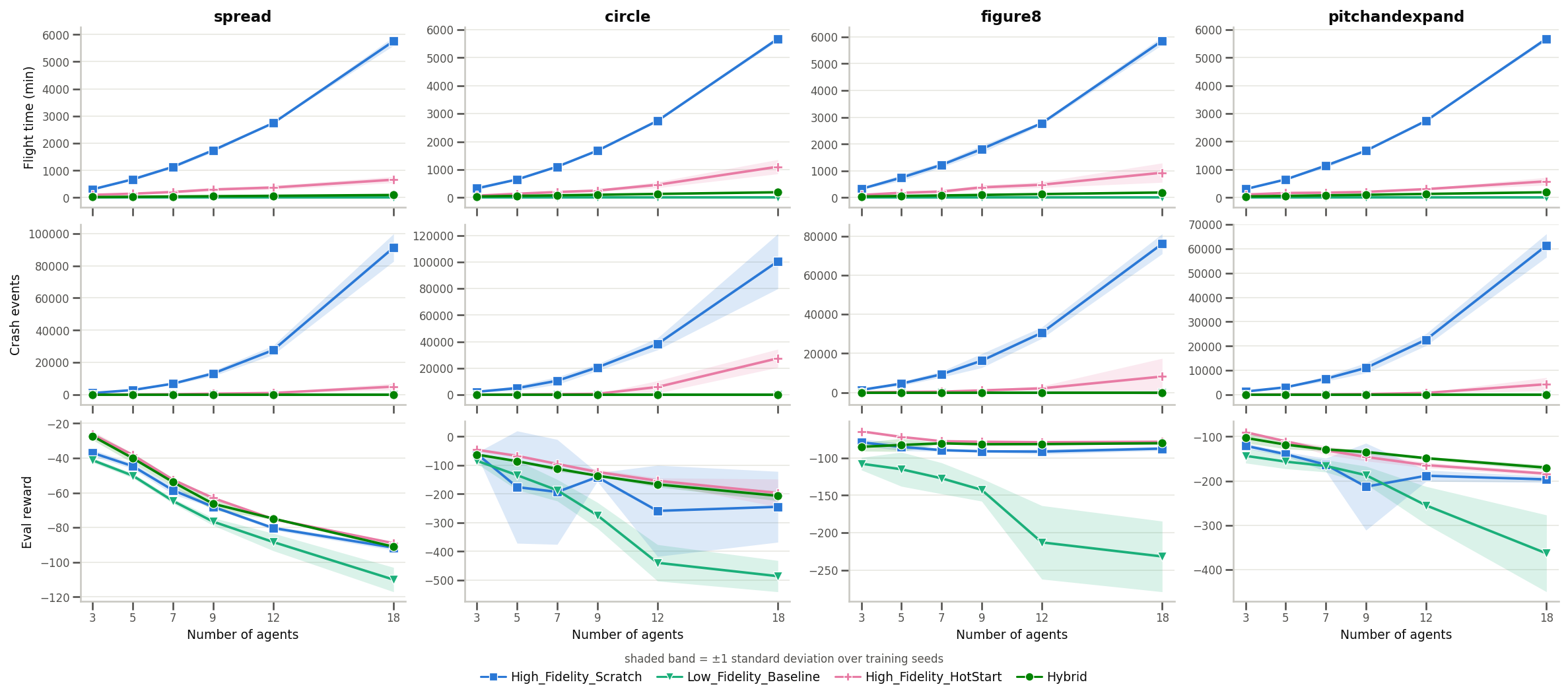}
\caption{Team-size scaling across all four tasks ($N\in\{3,5,7,9,12,18\}$) for the four
configurations of Section~\ref{sec:comparisons} (Hybrid is ours). Top: HF-simulation training
time; middle: HF-simulation training crash events; bottom: deployed evaluation reward
(mean $\pm$ std).}
\label{fig:summary_grid}
\end{figure*}

We evaluate the pipeline on four cooperative drone tasks (Figure~\ref{fig:scenario_trajectories}). 
These tasks are implemented identically across both pmDrones and crazyFlow, differing
only in physics fidelity, with all agents sharing a single policy. Every task also
adds a per-collision penalty to its reward:
\begin{itemize}
    \item \textbf{Spread:} Each agent is assigned a unique landmark on a randomly rotated formation 
    circle per episode. Agents are rewarded for proximity to their assigned landmarks and for 
    achieving full formation coverage as a team. This task uses the specialized near-hover
    loss reweighting during the residual fit (Section~\ref{sec:residual}).
    \item \textbf{Circle:} Agents must match a target tangential velocity while maintaining a 
    fixed-radius orbit at a specified altitude.
    \item \textbf{Figure8:} The entire swarm tracks a single moving target along a lemniscate 
    trajectory. This shared objective encourages collective movement, while the collision penalty 
    prevents spatial overlap.
    \item \textbf{Pitchandexpand:} Agents track a spherical shell centered on a static origin, 
    with a radius that oscillates sinusoidally over time. The reward function encourages both radial 
    proximity to the shell and matching the shell's purely radial velocity (via a bounded 
    $\exp(-\lVert\mathbf{v}-\mathbf{v}_{\mathrm{ideal}}\rVert)-1$ term) to penalize drifting across 
    the surface.
\end{itemize}

\subsection{Comparisons}
\label{sec:comparisons}

We evaluate four training and deployment configurations on crazyFlow (HF) rollouts:
\begin{itemize}
    \item \textbf{Low-Fidelity Baseline:} Trained entirely in LF physics without a residual or HF 
    involvement, isolating the contribution of the residual correction.
    \item \textbf{High-Fidelity Scratch:} Trained and evaluated entirely in HF physics from random 
    initialization, serving as an accurate but computationally expensive baseline.
    \item \textbf{High-Fidelity HotStart:} The Low-Fidelity Baseline fine-tuned briefly within 
    crazyFlow, providing a stronger HF reference.
    \item \textbf{Hybrid (ours):} The proposed pipeline (Section~\ref{sec:cheap-pipeline}): trained 
    in LF physics under a frozen, offline-fit residual ensemble, entirely avoiding HF reinforcement learning.
\end{itemize}

All configurations share the Hybrid's BPTT optimizer and policy architecture. Both HF baselines
run in a single crazyFlow world, so their reported HF costs reflect one physical drone
team, and both use a 100-epoch early-stopping patience. High-Fidelity Scratch gets a
linear-in-$N$ epoch budget large enough to converge ($\approx$1.9k epochs at $N{=}18$).
High-Fidelity HotStart, fine-tuning the converged LF policy at a reduced learning rate, needs far
fewer: across all 24 combinations and 5 seeds it early-stops well under its 500-epoch cap
(115--335 epochs).

To confirm the budget is not the limiting factor: despite $3$--$10\times$ more epochs (median
$\approx6\times$), Scratch never exceeds HotStart's training reward in any of the 24 cells (e.g.\
\textit{spread}, $N{=}12$: HotStart plateaus by epoch 185 above what a 1400-epoch Scratch run
reaches).

\section{Results}
\label{sec:results}

We evaluate sim-to-sim transfer, deploying each policy from pmDrones (LF) onto
crazyFlow (HF). To mirror real deployment, evaluation uses a single crazyFlow
world, not a batched ensemble. For every (scenario, team size) we report deployed episode
reward and, for the configurations that train in HF, their HF flight-time and crash cost. A crash
is registered per agent at the onset of contact, either inter-agent (center-to-center distance
below twice the collision radius) or with the floor, so that a sustained contact counts once and
$k$ agents colliding at the same step count as $k$. Floor contact in the first $0.5$\,s after a
reset is ignored, since every scenario spawns at ground level and liftoff takes $\approx0.3$\,s.
HF flight time is the simulated flight duration multiplied by the team size $N$ (fleet-seconds),
the airtime that would accrue across $N$ physical drones flying together. Each policy is evaluated
over $32$ episodes with independent initial conditions, and every training configuration is
trained for $5$ random seeds.

\subsection{Deployed Reward Across Team Sizes}
Figure~\ref{fig:summary_grid} evaluates all four configurations across the 24 conditions
formed by four scenarios and six team sizes ($N\in\{3,5,7,9,12,18\}$). Each setting is calibrated
from a pool of $60(N{+}1)$ single-drone reference trajectories of $10$\,s duration, scaling from
$240$ trajectories at $N{=}3$ to $1140$ at $N{=}18$. These reference flights are distributed
evenly across the $N$ team roles via deterministic round-robin assignment
(Section~\ref{sec:cheap-pipeline}); Section~\ref{sec:traj-sweep} analyzes the minimal calibration
budget required for convergence.

Hybrid outperforms the uncorrected Low-Fidelity Baseline in all 24 conditions, showing that a
residual fit purely to single-drone PD tracking data substantially bridges the sim-to-sim gap.
It also exceeds High-Fidelity Scratch in 22 of 24 settings: the from-scratch HF policy wins only
at $N{=}3$ on \textit{circle} and \textit{figure8}; at every larger $N\ge5$ Hybrid attains
higher rewards, often by wide margins (Fig.~\ref{fig:summary_grid}, bottom row).

Against High-Fidelity HotStart, the strongest baseline, and the only one trained directly on
the target dynamics, Hybrid trails at small team sizes, but the margin narrows steadily
with scale, from $\approx22\%$ at $N{=}3$ (mean over tasks) to under $1\%$ at $N{\ge}12$, where Hybrid matches
HotStart on \textit{spread} and exceeds it on \textit{pitchandexpand}. Team size is the
governing factor: both HF-trained baselines degrade sharply toward $N{=}18$, while Hybrid
degrades only mildly.

This performance divergence tracks training-time collision dynamics. Running BPTT directly in HF physics forces the policy through inter-agent contacts, and crash
frequency grows with swarm size (Fig.~\ref{fig:summary_grid}, middle row): at $N{=}18$ Scratch
incurs $6$--$10\times10^{4}$ crash events per run, a $16$--$63\%$ crash-step rate that biases
optimization toward sub-optimal minima (most visibly on \textit{pitchandexpand} at $N\ge9$ where
agents learn overly conservative collision-avoidance reflexes), and HotStart still shows $2$--$15\%$.
Because Hybrid optimizes entirely in LF physics, it is structurally immune to these contact
artifacts, incurring zero training crashes at every team size.

HF exposure shows the same disparity (Fig.~\ref{fig:summary_grid}, top row). Hybrid's HF
footprint is only Stage-2 calibration: $20$ fleet-minutes at $N{=}3$, $190$ at $N{=}18$, flown
by a deterministic controller on isolated vehicles rather than exploratory policies. HotStart
needs $70$--$1100$ fleet-minutes and Scratch $300$--$5900$. While these baseline costs grow
rapidly with $N$, Hybrid's calibration budget stays bounded and user-defined, and
Section~\ref{sec:traj-sweep} reduces it further.

\subsection{Calibration-Data Efficiency}
\label{sec:traj-sweep}

\begin{figure}[]
\centering
\includegraphics[width=\columnwidth]{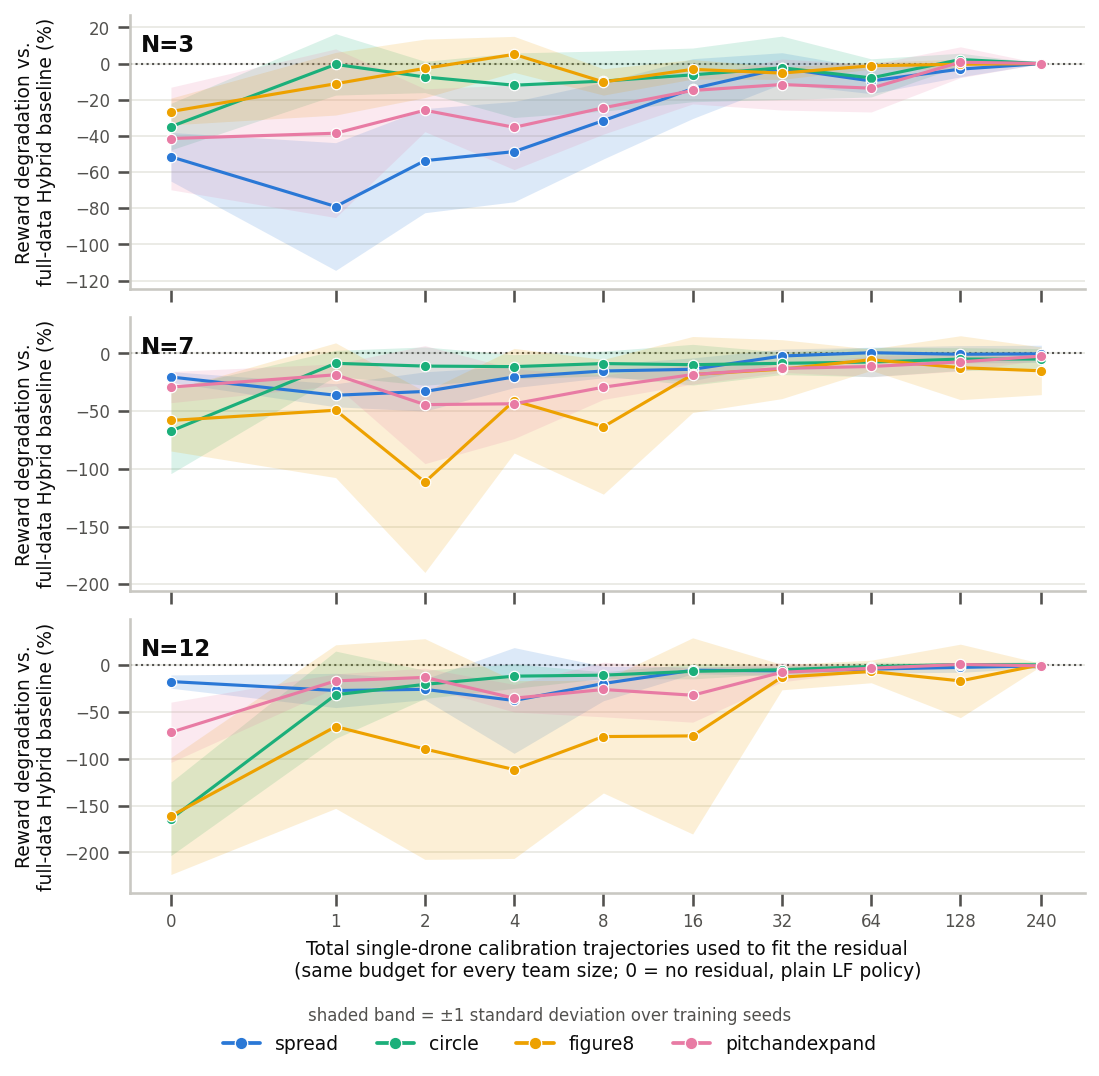}
\caption{Deployed reward vs.\ total HF calibration budget. One panel per team size
($N\in\{3,7,12\}$), one line per scenario; band is $\pm1$ std over training seeds. $y$: deployed
reward relative to that (scenario, $N$)'s own full-calibration-pool Hybrid
(Section~\ref{sec:comparisons}) (\%; $0=$
matched, negative $=$ worse). $x$: the \emph{total} number of single-drone PD-controller
calibration trajectories used to fit the residual ensemble, not per agent, and the same budget
regardless of team size (log scale; $x{=}0$ is the uncorrected LF policy).}
\label{fig:trajsweep_grid}
\end{figure}

The evaluations in Section~\ref{sec:comparisons} applied a fixed per-role trajectory budget. 
To evaluate calibration data requirements, we vary this budget from $0$ (no residual correction)
to $240$ trajectories, retraining the residual ensemble and Hybrid policy across five seeds
(Figure~\ref{fig:trajsweep_grid}). This analysis requires no additional simulations, as each condition 
subsamples the trajectory pool collected in Section~\ref{sec:cheap-pipeline}.

Deployed reward increases rapidly with initial calibration data before saturating. Without 
residual correction ($x{=}0$), the uncorrected LF policy trails the full-data Hybrid by $18$--$164\%$.
This gap expands with team size, reaching shortfalls of $-161\%$ in \textit{figure8} and $-164\%$
in \textit{circle} at $N{=}12$.

Calibration sensitivity differs across tasks:
\begin{itemize}
    \item \textit{Circle} converges fastest, reaching within roughly ten percent of the full-data fit by about $8$ calibration trajectories at every team size (a single trajectory already suffices at $N{=}3$).
    \item \textit{Spread} and \textit{Pitchandexpand} plateau by approximately $32$ trajectories.
    \item \textit{Figure8} requires the largest dataset. Below $32$ trajectories at $N\ge7$, cross-seed spread spans $30$--$110$ percentage points, narrowing once the budget reaches $32$--$64$ trajectories.
\end{itemize}

Across all four tasks and tested team sizes ($N\in\{3,7,12\}$), approximately $32$ trajectories, about
five fleet-minutes of single-drone flight, recover deployed performance to within roughly $10$--$15\%$ of the
full-pool baseline, with $64$ trajectories further reducing variance. This saturation threshold
remains independent of team size. While larger swarms exhibit higher variance under limited 
data due to covering broader regions of the state space, the total single-drone calibration 
volume required for convergence does not compound with $N$.

\subsection{Out-of-Distribution Deployment}
\label{sec:trajfollower}

\begin{figure}[]
\centering
\includegraphics[width=\columnwidth]{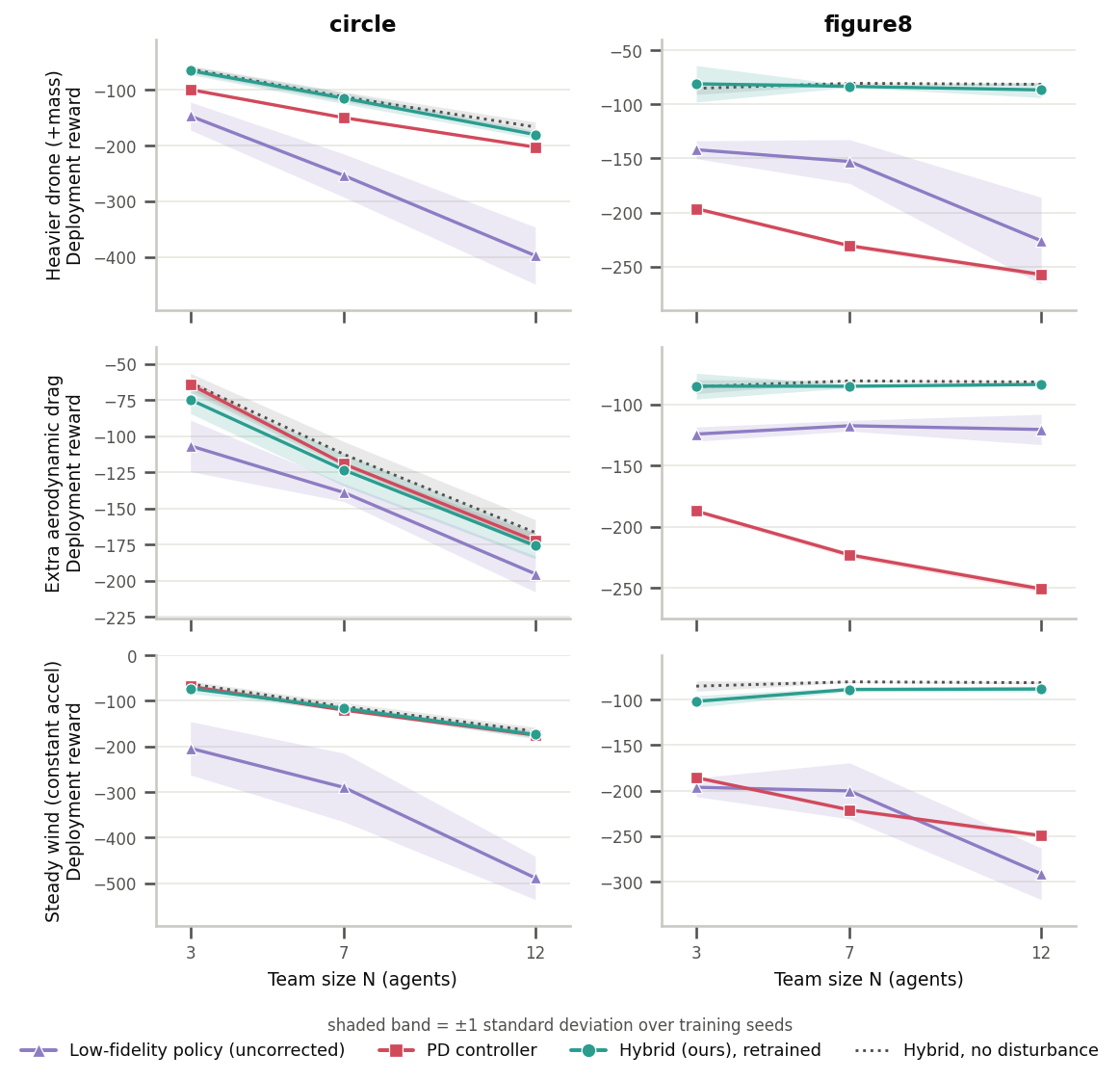}
\caption{Deployment on an off-spec HF drone. Rows: the three disturbances (heavier drone, excess
drag, steady wind); columns: scenarios; $x$: team size. Series: uncorrected LF policy deployed
directly, zero-training PD controller tracking the LF-policy reference, and Hybrid after a small
warm-started recalibration on the disturbed drone; dotted grey is Hybrid's clean-conditions
reward. Band is $\pm1$ std over $5$ training seeds.}
\label{fig:trajfollower_grid}
\end{figure}

The pipeline so far assumes calibration and deployment occur on an identical nominal model. To
test robustness under mismatch, we examine whether the residual-grounding stage can re-adapt the
policy to an off-spec platform. We use \textit{circle} and \textit{figure8}: both keep agents at
non-zero velocity throughout, so velocity-coupled and persistent-force disturbances are fully
engaged (unlike in station-keeping tasks), and they span distinct coordination
structures: a shared target vs.\ distributed formation control.

We evaluate three disturbances applied across both calibration and evaluation within crazyFlow 
at team sizes $N\in\{3,7,12\}$:
\begin{itemize}
    \item \textbf{Heavier drone} ($+20\%$ mass): Rescales the required thrust profile.
    \item \textbf{Excess aerodynamic drag} ($c = 2.0$\,s$^{-1}$): Introduces an unmodeled velocity-coupled drag force, $\mathbf{f} = -c\,m\,\mathbf{v}$.
    \item \textbf{Steady wind} ($1.5$\,m/s$^2$ bias): Applies a persistent directional acceleration to each agent.
\end{itemize}

The uncorrected LF baseline and the PD controller are deployed unmodified. For Hybrid we run a
brief warm-started recalibration from its clean checkpoints: 20 PD calibration trajectories on
the disturbed model, 150 epochs of residual refit, and 400 BPTT epochs of policy fine-tuning
under the re-fit residual. All configurations are then evaluated on the disturbed HF system.

On \textit{figure8}, the recalibrated Hybrid shows strong resilience, maintaining returns between 
$-81$ and $-102$ across all disturbances and swarm sizes, closely matching its clean baseline of
approximately $-82$. In contrast, the uncorrected policy drops to between $-120$ and $-292$. 
On \textit{circle}, adaptation provides substantial recovery, though performance depends more on the 
underlying disturbance:
\begin{itemize}
    \item Under the heavier mass profile, the uncorrected baseline drops from $-147$ to $-398$ as team size scales. The recalibrated Hybrid maintains returns between $-66$ and $-180$, tracking its clean baseline range of $-63$ to $-167$.
    \item Under steady wind, the uncorrected policy degrades to $-204$ through $-489$, whereas Hybrid stabilizes between $-73$ and $-174$.
    \item Under excess drag, performance degradation is less severe overall; Hybrid's margin over the PD tracking baseline narrows, but it reliably outperforms the uncorrected policy.
\end{itemize}

Although the PD controller tracks the agile \textit{figure8} reference loosely (trailing even the
uncorrected policy), this does not undermine calibration: the residual is fit from observed
transition pairs, not tracking precision. Overall, a few minutes of single-drone recalibration
lets the frozen residual compensate for substantial platform mismatch, and because that
recalibration uses only local tracking transitions, its cost stays decoupled from swarm size.

\subsection{Residual Model Class}
\label{sec:residual-class}

\begin{table}[H]
\centering
\resizebox{\columnwidth}{!}{%
\begin{tabular}{l|ccc}
\hline
\textbf{Task} ($N$) & \textbf{Affine} & \textbf{SINDy} & \textbf{MLP ens.} \\
\hline
Spread ($N{=}3$)          & $-32.9{\pm}6.0$    & $-28.1{\pm}0.4$          & $\mathbf{-27.7{\pm}2.7}$ \\
Spread ($N{=}12$)         & $-88.3{\pm}11.7$   & $\mathbf{-77.1{\pm}0.5}$ & $-77.5{\pm}3.7$ \\
Circle ($N{=}3$)          & $-289.6{\pm}220.1$ & $\mathbf{-48.0{\pm}1.8}$ & $-61.5{\pm}4.1$ \\
Circle ($N{=}12$)         & $-276.7{\pm}165.6$ & $-287.3{\pm}63.2$        & $\mathbf{-168.3{\pm}8.1}$ \\
Figure8 ($N{=}3$)         & $-91.0{\pm}2.8$    & $\mathbf{-76.9{\pm}2.3}$ & $-85.2{\pm}4.5$ \\
Figure8 ($N{=}12$)        & $-85.6{\pm}1.0$    & $-117.5{\pm}20.3$        & $\mathbf{-81.1{\pm}0.7}$ \\
Pitchandexpand ($N{=}3$)  & $-255.4{\pm}247.2$ & $\mathbf{-90.8{\pm}3.8}$ & $-106.0{\pm}12.1$ \\
Pitchandexpand ($N{=}12$) & $-173.3{\pm}15.5$  & $-159.9{\pm}9.7$         & $\mathbf{-151.9{\pm}7.2}$ \\
\hline
\end{tabular}%
}
\caption{Deployed reward (mean\,$\pm$\,std over five seeds) across residual classes. Bold indicates best mean per row.}
\label{tab:residual-class}
\end{table}

To assess sensitivity to the residual's parameterization, we evaluate three frozen model classes under an identical pipeline: (i) an affine correction, (ii) a SINDy model \cite{bruntonDiscoveringGoverningEquations2016} fit by sequentially-thresholded least squares over a physically grounded library (thrust, attitude projections, linear velocity), and (iii) the bootstrap-bagged MLP ensemble. All share the same bounded output; the affine and MLP variants are five-member bagged ensembles averaged uniformly, while the SINDy model is a single sparse fit.

At $N{=}3$, performance is comparable across classes, with SINDy achieving the highest mean return. However, stability diverges: the affine model collapses on \textit{circle} and \textit{pitchandexpand} at $N{=}3$ ($\mathrm{std} > 200$), while SINDy degrades at $N{=}12$ on \textit{circle} and \textit{figure8}. The MLP ensemble is the only class that avoids divergence across all settings, making it our default; the pipeline itself remains model-agnostic. That the simpler classes stay competitive at all partly reflects crazyFlow's deterministic, near-affine gap; a stochastic or more strongly nonlinear gap would widen the margin in favor of the expressive ensemble.

\section{Conclusion}
\label{sec:conclusion}
We bridge the sim-to-sim fidelity gap for cooperative drone swarms using an offline-fit,
frozen residual acceleration model. A bootstrap-bagged MLP ensemble fit to short,
single-drone PD tracking flights allows the cooperative policy to be optimized entirely
within nominal point-mass physics, circumventing multi-agent HF reinforcement learning.
Because the dynamics correction operates strictly per-agent, swarms across all tested
team sizes are grounded and trained without requiring a single multi-drone flight.

Across four cooperative tasks and team sizes $N\in\{3,5,7,9,12,18\}$, the proposed hybrid
pipeline outperforms an uncorrected LF baseline in all 24 configurations and exceeds a
from-scratch HF baseline in 22 of 24 (trailing only at $N{=}3$). Its performance gap
relative to an HF fine-tuning baseline narrows steadily from $\approx 22\%$ at
$N{=}3$ to under $1\%$ at $N{\ge}12$, while consuming a fraction of the HF simulation time.
Furthermore, while multi-agent contact dynamics cause training crash-step rates in HF
baselines to exceed $60\%$ at $N{=}18$, our LF optimization avoids physical collision
instabilities entirely, maintaining zero training crashes. The total HF requirement is
restricted to an initial calibration phase that can be compressed to roughly $32$
single-drone trajectories (about five fleet-minutes) without substantial performance loss.

Several directions remain. All evaluations are in simulation; although crazyFlow is
system-identified against physical hardware, real-world validation is an essential next step.
crazyFlow is also deterministic, without sensor or actuation noise; evaluating the
framework under stochastic dynamics and coupled, multi-axis disturbances warrants further study.
Still, the methodology is agnostic to aerial robotics, needing only an analytical nominal model,
a high-fidelity reference, and a basic tracking controller, and this per-agent residual paradigm
is a scalable foundation for other multi-agent systems, including ground rovers and autonomous
surface vessels.

\bibliographystyle{ieeetr}
\bibliography{references}

\end{document}